\documentclass{aa}
\usepackage{graphicx}
\begin{document}
\title{A high-resolution study of the evolution of the \\
Lyman $\alpha$ forest in the redshift interval $0.9 < z < 1.7$ 
\thanks{Based on observations with the NASA/ESA Hubble Space
Telescope, obtained at the Space Telescope Science Institute, which is 
operated by Aura, Inc., under NASA contract NAS 5-26555; and on observations
made with ESO telescopes at the La Silla or Paranal Observatories under  
programme ID 066.A-0212}}
 
\author{E.~Janknecht, R.~Baade, and D.~Reimers}
\authorrunning{E. Janknecht et al.}   
\titlerunning{First high-resolution study of the Lyman $\alpha$ Forest 
at $0.9 < z < 1.7$}
\institute{Hamburger Sternwarte, Universit\"at Hamburg, Gojenbergsweg 112,
D-21029 Hamburg, Germany\\
\email{[ejanknecht,rbaade,dreimers]@hs.uni-hamburg.de}}

\offprints{E.~Janknecht,\\
\email{ejanknecht@hs.uni-hamburg.de}}

\date{Received \today / Accepted}

\abstract{

Spectroscopy with {\it HST/STIS} Echelle and {\it VLT/UVES} 
of the bright QSO HE 0515$-$4414 ($z_{\rm em} =1.73, B = 15.0$) 
offers for the first time the opportunity to study the Lyman $\alpha$ forest
in the redshift range $0.9 < z < 1.7$ at a resolution $\leq 10$ km\,s$^{-1}$. 
The number density evolution of the Lyman $\alpha$ lines is well described
by the power law approach d$n/{\rm d}z \propto (1+z)^\gamma$. We derive
$\gamma = 2.23 \pm 1.21$ for the strong lines 
($13.64 \leq  {\rm log}\, N_{\rm \ion{H}{i}} \leq 16.00$),
in agreement with the Lyman $\alpha$ forest evolution for $z > 1.7$. 
The expected slow-down in their evolution does not appear 
earlier than $z \sim 1$. For the weak lines 
($13.10 \leq  {\rm log}\, N_{\rm \ion{H}{i}} \leq 14.00$)  we find 
that the HE 0515$-$4414 data for $z > 1$ follow the trend with $\gamma = 1.40$
known from $z > 1.7$ observations, i.e. we confirm the difference in evolution
between weak and strong lines. 
We use the two-point velocity correlation function (TPCF) 
to search for clustering of the Lyman $\alpha$ lines, 
yet we detect no excess in the TPCF on scales up to 10\,000 km\,s$^{-1}$.

}

\maketitle

\section{Introduction}

The evolution of the Lyman $\alpha$ absorption lines per unit redshift 
d$n/{\rm d}z$ 
has been a subject of observational studies since many years. At high redshift 
($z>1.6$) the evolution is steep with d$n$/d$z \propto (1+z)^{\gamma}$
and $\gamma = 2 - 3$ for strong Lyman $\alpha$ lines (i.e. 
log $N_{\rm \ion{H}{i}} > 14$; \cite{Rauch}).
Observations of the Lyman $\alpha$ forest at low redshift with {\it HST/FOS}, 
at first in 3C273 (\cite{Bahcall}), showed that the number of 
Lyman $\alpha$ lines was far in excess of the expected number according 
to an extrapolation from high $z$. As a final result from the HST QSO 
absorption line key project \cite{Weymann()}\, found $\gamma = 0.1 - 0.3$ 
for $z < 1.5$. 
They also claimed a break in the evolutionary behaviour from 
a steep evolutionary law for $z > 1.6$ to a flat one for $z < 1.5$. 
The apparently rather abrupt break in the evolutionary law 
just at the transition from high-resolution optical data to low-resolution 
UV data immediately raised the suspicion that this behaviour might 
not be real. Owing to the insufficient FOS resolution 
($230 - 270 \ {\rm km\,s}^{-1}$) the number counts might be underestimated 
due to line blending (\cite{Weymann}). It is obvious, that this open 
question can be addressed only by UV observations of very bright QSOs at 
Echelle resolution. A further unsettled question which requires 
high-resolution UV spectra is the different evolutionary behaviour of strong 
compared to weak Lyman $\alpha$ lines. It has been found from high- 
resolution optical spectra for $z > 1.6$ that apparently weak lines evolve 
much slower than strong lines (\cite{Kim1}). Is that also true 
for $z < 1.5$? 

In this paper we use the first UV spectra of the bright 
intermediate redshift QSO HE 0515$-$4414 ($z = 1.73$, $B = 15.0$, 
\cite{Reimers}) at {\it STIS}/Echelle resolution (10 km\,s$^{-1}$)
for addressing the above discussed open questions.

\section{Observations}

HE\,0515$-$4414 was observed with {\it STIS} between January 31 
and February 2, 2000 with the medium-resolution {\it NUV} echelle mode (E230M) 
and a 0.2 x 0.2 aperture. The overall exposure time was 31\,500 s
resulting in a typical signal-to-noise ratio of 10 depending on the order 
and on the position within the orders. The resolution of the spectra is 
FWHM$\, \simeq$ 10 km\,s$^{-1}$.
The data reduction was performed by the {\it HST} pipeline completed by
an additional inter-order background correction and by coadding the separate
subexposures. 
 
The optical spectra were obtained between October 7, 2000 and January 3, 2001
using the {\it UV-Visual Echelle Spectrograph (UVES)} at the {\it VLT}/Kueyen 
telescope. The overall exposure time was 31\,500 s. The slit width was 0.8
arcsec resulting in a spectral resolution of 
FWHM$\,\simeq$~6~km\,s$^{-1}$. After reduction by the {\it UVES} pipeline
and conversion to vacuum baryocentric wavelengths, the individual spectra
were coadded and exhibit a S/N $\simeq 10 - 50$ in the investigated 
spectral region.

\section{Data analysis}

The combined {\it HST} and {\it VLT} data provide the spectral range
of the Lyman $\alpha$ forest of HE\,0515$-$4414 from $z=0.87$ up to $z=1.73$,
the quasar's Lyman $\alpha$ emission redshift. 
To avoid the proximity effect we exclude a region of about 5000 km\,s$^{-1}$
from the quasar leading to an investigated spectral range 
$\lambda = 2278 - 3260$ \,\AA\, or $z = 0.87 - 1.68$. 
We normalize the spectrum fitting polynomials to line-free regions and
dividing the flux by this background continuum.

The spectrum of HE\,0515$-$4414 is strewn with metal lines and lines of molecular
hydrogen (H$_{2}$) from its damped Lyman $\alpha$ system which has been
studied in detail by \cite{Ana}. The main difficulty 
in the line identification was the extraction of the 
Lyman $\alpha$ lines from this real H$_{2}$ 
forest dominating the short-wavelength region of the {\it HST} spectrum
and thus suggesting to ignore this region in our analysis of the weak Lyman 
$\alpha$ lines (see below). The molecular hydrogen of the DLA 
will be the topic of a forthcoming paper.  

In a first approximation, we detected about 400 lines as Lyman $\alpha$ 
candidates in the whole spectrum. These lines were fitted with the FITLYMAN 
code in the MIDAS package (\cite{Fontana}) using Voigt profiles 
convolved with the instrumental profile. FITLYMAN adjusts three independent
parameters per line by $\chi^{2}$ minimization.
These parameters are the redshift of an absorption line $z$, 
its \ion{H}{i} column density $N_{\rm \ion{H}{i}}$, and its Doppler 
parameter $b$, 
comprising the thermal and the turbulent broadening of the lines.
The general fitting strategy, especially for blends, was to start with a 
single line and to add a further component to the ensemble if the $\chi^{2}$ 
decreased with this second line. 
For each of the Lyman $\alpha$ candidates we calculated the significance level
$S = \frac {W}{\sigma_{W}}$, where $W$ denotes the observed equivalent width 
of the line and $\sigma_{W}$ the $1\, \sigma$ error of $W$ implying both
the fit error and the continuum error. 
With the selection criteria $S \ga 1$ and $b \ga 10$ km\,s$^{-1}$,
we reduced the original sample to 235 
Lyman $\alpha$ lines. The full fit parameter list for all recognized hydrogen 
absorption lines is available in electronic form at the CDS 
via anonymous ftp to cdsarc.u-strasbg.fr.

\section{Discussion}

\subsection{Simultaneous fits with Lyman $\beta$}

An important fitting constraint is given for the stronger Lyman $\alpha$ 
lines for which we tried to detect the higher order lines of the Lyman series
followed by a simultaneous fit, if possible. We compared the \ion{H}{i} 
parameters inferred from Ly $\alpha$ profile fitting with the parameters
determined with a simultaneous fit with the accompanying Ly $\beta$ 
counterparts. Fig. 1 shows a rather small systematic effect 
in the column densities. We derive 
$\langle\, N_{\rm \ion{H}{i}}^{\rm sim} / N_{\rm \ion{H}{i}}^{\rm single} 
\rangle = 1.145$ 
for the mean proportion of the Ly $\alpha$/Ly $\beta$ fit column densities 
and the column densities based solely on Ly $\alpha$ fitting.
The same holds for $b$  
($\langle\,b_{\rm sim} / b_{\rm single}\rangle = 1.03$).
This is in contrast to the recently suggested strong dependence of the fit 
parameters on the fit strategy (single-component fit versus simultaneous 
Ly $\alpha$-Ly $\beta$ fit; e.g., \cite{Hurwitz}; \cite{Shull}). 
Our 15 per cent deviation of the inferred column densities is at variance
with the above cited studies. Obviously, the improved data quality 
leads to more consistent fit results. The remaining discrepancy may be
attributed to the unphysical assumption of Voigt profile 
fitting. Indeed, the interpretation of the line-broadening velocity 
as unresolved stochastic motions is probably an oversimplification 
(e.g., \cite{Levshakov}).

\begin{figure}
\hspace*{1cm}
\includegraphics[width=6cm,angle=90]{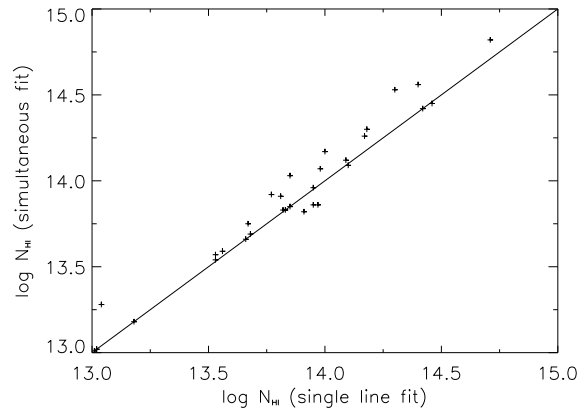}
\caption{Column densities of individually treated Lyman 
$\alpha$ lines versus column densities of 
simultaneous  Ly $\alpha$/Ly $\beta$ fits. We present the results 
for all detected pairs.}
\label{alpha-beta}
\end{figure}

\subsection{Column density distribution function}

The differential column density distribution function 
$f(N_{\rm \ion{H}{i}})$ 
is defined as the number of Lyman $\alpha$ absorption lines per unit
column density and per unit absorption distance path (\cite{Tytler}):

\begin{equation}
f(N_{\rm \ion{H}{i}}) = \frac {n}{\Delta N \, \Sigma_{i}\Delta X_{i}}
\end{equation}

\noindent
For $q_{0} = 0$, $X(z) = \frac {1}{2}\,\, [(1+z)^{2} - 1]$, 
so that 
$\Delta X = (1~+~z)\, \Delta z$. For the spectral range investigated we get 
$\Delta z = 0.808$, $z = \overline{z} = 1.278$, and hence 
$\Delta X = 1.841$.

\begin{figure}
\hspace*{-0.8cm}
\includegraphics[width=9.3cm]{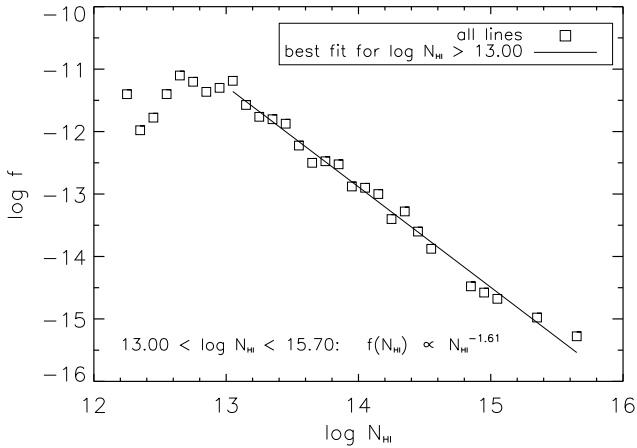}
\caption{The column density distribution function for the Lyman $\alpha$ lines.
For the best fit only lines with 
log ${\sl N}_{\rm \ion{H}{i}} > 13.00$ were included
due to the limited S/N of the spectra.}
\label{CDDF} 
\end{figure}

Usually, $f(N_{\rm \ion{H}{i}})$ is fitted by a power law of the form
$f(N_{\rm \ion{H}{i}}) = A\, N_{\rm \ion{H}{i}}^{-{\beta}}$. 
The distribution function for all 232 lines with 
log $N_{\rm \ion{H}{i}} \leq 15.70$ 
is plotted in Fig. 2. The squares show the observed log $f$ values, 
while the solid line represents the best fit for all lines with
log $N_{\rm \ion{H}{i}} \geq 13.00$. We have chosen this lower boundary 
because the distribution follows the power law down 
to this value. Assuming the general validity of the power law the sample
is obviously not complete below log $N_{\rm \ion{H}{i}} \la 13.00$. 
  
The best fit yields log $A = 9.62\, \pm\, 0.58$ and $\beta = 1.61 \pm 0.04$.
Considering exclusively the {\it STIS} lines in the same column density 
range, the result is very similar: log $A = 9.47 \pm 0.59$ and 
$\beta = 1.60 \pm 0.04$. In contrast, the slope   
of the distribution function for the {\it UVES}\,\, lines is much flatter 
(log $A = 4.60 \,\pm\, 0.72$, $\beta = 1.24 \pm 0.05$).
The distribution of the {\it UVES} lines can be approximated very well
with a single power law for an even wider column density range 
($12.50  \leq  {\rm log}\, N_{\rm \ion{H}{i}} \leq 15.70$)
due to the better resolution of {\it UVES}.
The flatter slope of the higher redshifted {\it UVES} lines indicates 
that stronger absorbers have evolved away faster than weaker ones.
This will be discussed in more detail in Section 4.3.

Our result is in accordance with other analyses in comparable redshift ranges.
For example, \cite{Dobrzycki()} found 
$\beta \leq 1.6 - 1.7$, deriving the exponent from a curve of growth 
analysis. \cite{Hu}  obtained $\beta = 1.46$ for 
$12.3 \leq  {\rm log}\, N_{\rm \ion{H}{i}} \leq 14.5$, while 
\cite{Kim2()} determined $\beta$ for various column density ranges 
to $1.70-1.74$.

\subsection{Number density evolution}

The evolution of the number density per unit redshift of 
Lyman $\alpha$ clouds can be well approximated by the power law

\begin{equation}
\frac{\rm d \it n}{\rm d \it z} = \left(\frac{\rm d \it n}{\rm d \it z}
\right)_{0} \,\, (1 + z)^ {\gamma},
\end{equation}
  
\noindent
where $\left(\frac{\rm d \it n}{\rm d \it z}\right)_{0}$ is the local 
comoving number density and the exponent $\gamma$ includes the 
cosmological evolution as well as the intrinsic evolution of the absorbers. 
For a non-evolving population in the standard Friedmann universe 
with the cosmological constant $\Lambda = 0$ and with $q_{0} = 0$
the exponent becomes $\gamma = 1$. 
   
Because there exist a lot of hints that weaker and stronger 
Lyman $\alpha$ clouds evolve differently, it is convenient to distinguish
between two subsamples. We use the column density range
$13.10 \leq  {\rm log}\, N_{\rm \ion{H}{i}} \leq 14.00$ 
for the weak lines, while the interval 
$13.64 \leq  {\rm log}\, N_{\rm \ion{H}{i}} \leq 16.00$ 
defines the strong ones. These boundaries are frequently used in the
literature (\cite{Weymann}; \cite{Kim2}; \cite{Dobrzycki}) 
allowing a direct comparison.

\begin{figure}
\hspace*{-0.8cm}
\includegraphics[width=10cm]{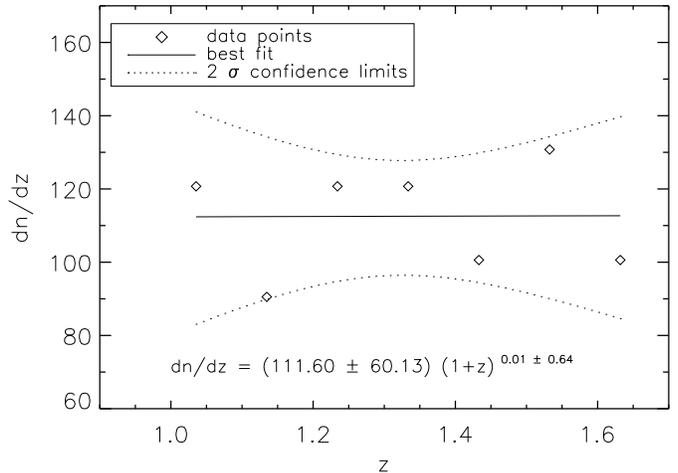}
\caption{The number density evolution of the Lyman $\alpha$ forest for weak 
lines ($13.10 < {\sl N}_{\rm \ion{H}{i}} < 14.00$). 
The data points are binned with 
$\Delta z = 0.1$. The best fit was obtained by $\chi^2$ minimization. The 
dotted curves represent the 95 \% confidence band.}
\label{weak}
\end{figure}

\begin{figure}
\hspace*{-0.8cm}
\includegraphics[width=10cm]{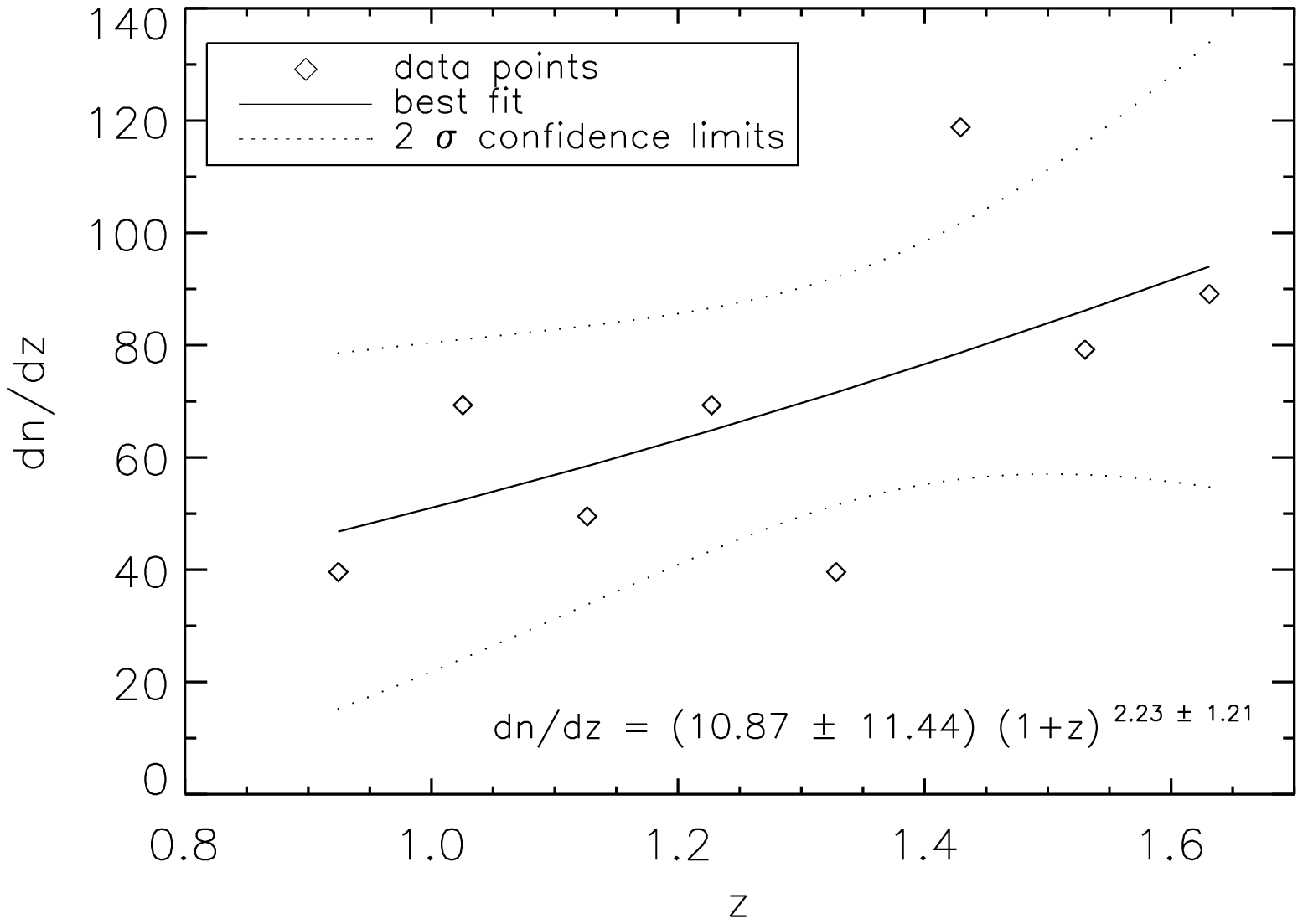}
\caption{Same as in Fig. 3, for $13.64 <  {\sl N}_{\rm \ion{H}{i}} < 16.00$.}
\label{strong} 
\end{figure}

In Fig. 3 and Fig. 4 we present the line numbers per unit redshift plotted
over the redshift for the weak and for the strong lines, respectively. 
While for the strong lines we could exploit the whole wavelength region, 
we omit the spectral range $\lambda \la 2415$ \,\AA\,
for the weak lines to avoid misidentification due to the H$_{2}$ lines, 
retaining an effective redshift range $1.0 \la z \la 1.7$. 
The diagrams show the data points and the best fit for which we obtain 
$\gamma = 0.01 \pm 0.64$ for $13.10 \leq  
{\rm log}\, N_{\rm \ion{H}{i}} \leq 14.00$,
suggesting that there is little
evolution in the weak lines in the redshift interval 
$1.0 < z < 1.7$. Considering a broader redshift range (Fig. 5, upper
part) demonstrates that our {\it STIS} data follow the earlier
optical observations ($z > 1.6$), i.e. the number density evolution is
consistent with $\gamma = 1.40$ over the whole redshift range 
$1 \leq z \leq 3.7$. It should be noted that $\gamma = 1.40$ lies 
within the 2$\sigma$ confidence band of our data points (see Fig. 3). 
The transition to a flat ($\gamma = 0$) evolution curve probably 
occurs around $z = 1$.   
     
In contrast, the strong Lyman $\alpha$ lines show a steeper gradient
in the evolution diagram (Fig. 4).  
We detect an obvious correlation between the evolution and the line strength, 
i.e., the high column density absorbers evolve with 
$\gamma = 2.23 \pm 1.21$. This disagrees with the results of \cite{Penton()} 
and \cite{Dobrzycki()} who found no or only marginal evidence 
for a different evolution, respectively.
    
We cannot recognize a slow-down in the evolution of the stronger absorbers.
Therefore, we conclude that this break does not occur earlier than at 
$z \sim 1$
rather than at $z \sim 1.5 - 1.7 $ as previously claimed
(\cite{Impey}; \cite{Weymann}; \cite{Dobrzycki}).  
The large spread of our data points result from the poor statistics of 
a single line of sight. 
For example, omitting the two outliers in the plot for the strong lines 
lying beyond the 95 \% confidence limit our result 
($\gamma = 1.88 \pm 0.57$) becomes more robust. 
  
The lower panel of Fig. 5 demonstrates the difference  
between the \cite{Weymann()} data points and our values. 
Indeed, the former ones do not indicate any change 
in the evolution until $z \sim 1.5$, while the results of HE 0515$-$4414 
suggest a change of the slope at much lower $z$.

\begin{figure}
\hspace*{-0.2cm}
\includegraphics[width=9cm]{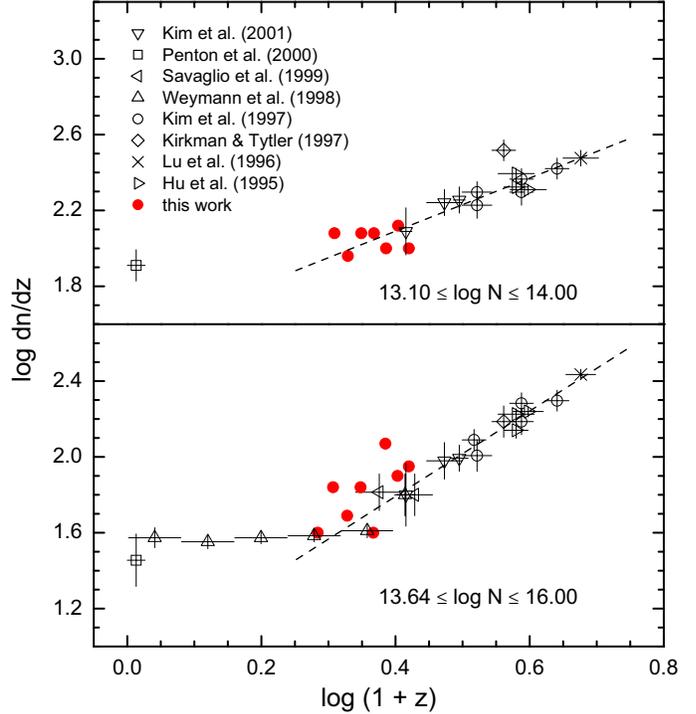}
\caption{The number density evolution of the Lyman $\alpha$ forest, 
comparison of different studies. Shown are the results for the column
density range ${\sl N}_{\rm \ion{H}{i}} = 
10^{13.10 - 14.00} {\rm cm}^{-2}$ (upper panel) and 
${\sl N}_{\rm \ion{H}{i}} = 10^{13.64 - 16.00} {\rm cm}^{-2}$ 
(lower panel). The filled circles are from our study, for the other symbols
see legend. It should be noted that the data points from 
\cite{Penton()} and \cite{Savaglio()} in the lower panel apply to 
${\sl N}_{\rm \ion{H}{i}} > 10^{14.00}{\rm cm}^{-2}$.}
\label{nde_Lit} 
\end{figure}

\subsection{Two-point velocity correlation function}

To study the clustering properties of the Lyman $\alpha$ forest,
we introduce the two-point velocity correlation function $\xi\, (\Delta v)$
where the number of observed line pairs in a given velocity 
separation bin $\Delta v$,\, $n_{\rm obs}$, is compared with the number of
expected pairs $n_{\rm exp}$ determined in the same velocity difference bin
in a randomly produced spectrum:
 
\begin{equation}
\xi\, (\Delta v) = \frac{n_{\rm obs}\, (\Delta v)}{n_{\rm exp}\, (\Delta v)} 
- 1.
\end{equation}

\noindent
Having two absorbers at $z_{1}$ and $z_{2}$ the velocity splitting 
in the rest frame at mean redshift reads

\begin{equation}
\Delta v =  \frac {c\,(z_{2} - z_{1})}{1 + \frac {z_{1} + z_{2}}{2}}.
\end{equation} 
 
\noindent
We derived $n_{\rm exp}$ from Monte Carlo simulations distributing our line
sample - in accordance with the determined number density evolution - 
randomly over the spectrum, counting the line pairs for varying 
velocity splittings and repeating this procedure 1000 times.    
 
Various studies have found at least marginal evidence for clustering 
on scales 
$\Delta v \leq 500\ {\rm km\,s}^{-1}$. For example, 
\cite{Cristiani()} derived   
$\xi = 0.89 \pm 0.18$ and $\xi = 1.02 \pm 0.26$ (for two different QSOs)
for ${\rm log}\, N_{\rm \ion{H}{i}} > 13.8$ and a redshift interval
 $0 < z < 3.66$ at a velocity separation of 
$\Delta v = 100\ {\rm km\,s}^{-1}$, while \cite{Kim2()} found
$\xi = 0.4 \pm 0.1$ for ${\rm log}\, N_{\rm \ion{H}{i}} > 12.7$
and $1.5 < z < 2.4$ at the same velocity splitting.
\cite{Penton()} and \cite{Kulkarni()} also detected 
a weak clustering signal in their data. 
In addition, the occurence of clustering on different scales is also 
supported by a number of theoretical studies
(e.g., \cite{Cirkovic}, \cite{Pando}). 
 
We have examined the clustering separately for the weak and strong lines
(the latter is shown in Fig.~6). 
Up to a velocity separation of $\Delta v = 1000\ {\rm km\,s}^{-1}$ 
we detect no signal exceeding the 2$\sigma$ level and only a marginal one
above the 1$\sigma$ level. For example, for the strong lines we derive 
$\xi\,\left(\overline{\Delta v}=50\, {\rm km\,s}^{-1}\right) = 1.14$
with a 1$\sigma$ poisson error of 1.13 in accordance with Fig. 6 . For 
${\rm log}\, N_{\rm \ion{H}{i}} > 13.8$ we obtain 
$\xi\, (50\, {\rm km\,s}^{-1}) = 1.63$ (1$\sigma$ poisson error 1.60)
and $\xi\, (150\, {\rm km\,s}^{-1}) = 1.23$ (1$\sigma$ poisson error 1.36),
so we cannot confirm the \cite{Cristiani()} result. 
For ${\rm log}\, N_{\rm \ion{H}{i}} > 12.7$,
$\xi\, (50\, {\rm km\,s}^{-1}) = 0.10$ (1$\sigma$ poisson error 0.16) and 
$\xi\, (150\, {\rm km\,s}^{-1}) = -0.01$ (1$\sigma$ poisson error 0.15)
not being consistent with the weak clustering seen by \cite{Kim2()} 
in a comparable redshift range.
There is also no excess in $\xi$ on much larger velocity separations 
(up to 10\,000 km\,s$^{-1}$) in our data. We do not find any 
dependence of $\xi$ on the redshift either. 
Thus we cannot confirm the result of weak clustering of 
the Lyman $\alpha$ lines as suggested in the above cited studies.
However, the small number of lines in a single line of sight     
does not allow a final assessment. 
   
\begin{figure}
\hspace*{-2cm}
\includegraphics[width=8cm,angle=90]{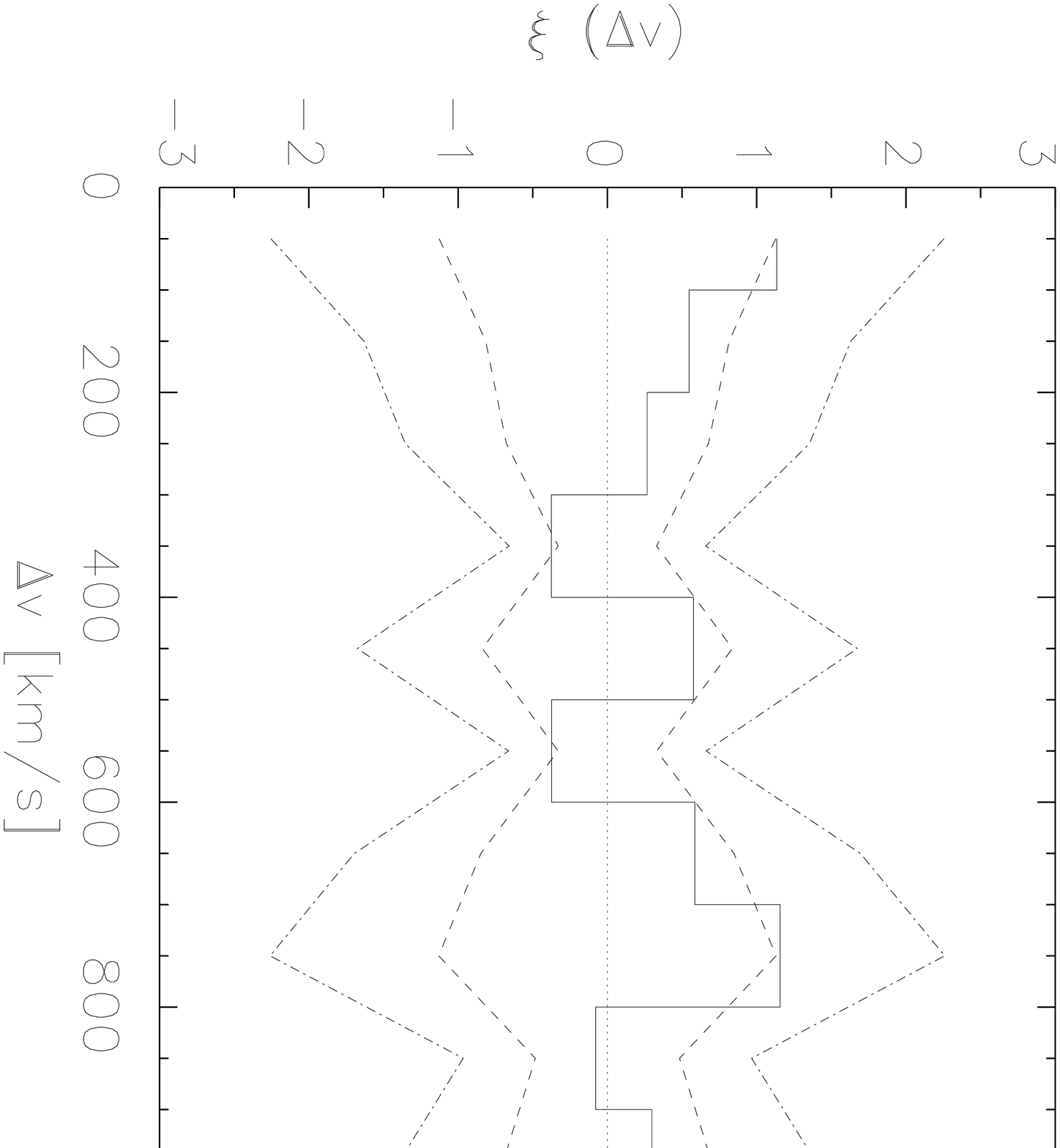}
\caption{Two-point correlation function $\xi\, (\Delta v)$ (solid line),
in 100 km\,s$^{-1}$ bins, for higher column density absorbers 
($13.64 \leq  {\rm log}\, N_{\rm HI} \leq 16.00$). Dashed and dot-dashed lines 
represent the 1$\sigma$ and 2$\sigma$ Poisson errors, respectively.}
\label{clust_strong}
\end{figure}

\section{Conclusions}

We have analyzed the evolution of the Lyman $\alpha$ forest in the 
redshift range $0.9 < z < 1.7$ using combined high-resolution {\it HST/STIS}
and {\it VLT/UVES} data. The main results are summarized as follows: 

\begin{enumerate}

\item    
 
The recently claimed dependence of the hydrogen fit parameters on the 
fit strategy (single Ly $\alpha$ fit or simultaneous Ly $\alpha$-Ly $\beta$ 
fit) cannot be confirmed by our results. 

\item    
 
The column density distribution of our Lyman $\alpha$ line sample can be 
approximated over several orders of magnitude 
($13.00 \leq  {\rm log}\, N_{\rm \ion{H}{i}} \leq 15.70$)
by a power law. The (negative) slope is described by 
$\beta = 1.61\, \pm\, 0.04$ \,for $\overline{z} \sim  1.3$, 
consistent with recent results. 
 
\item    

The evolution of strong and weak lines is distinctly different. 
The high column density 
($13.64 \leq  {\rm log}\, N_{\rm \ion{H}{i}} \leq 16.00$) 
absorbers evolve according to $(1+z)^{\gamma}$ with
$\gamma = 2.23\, \pm\,1.21$ for $0.9 < z < 1.7$, and the 
expected slow-down in the evolution does not appear down to $z \sim 1$.
The evolution of the weaker lines over the same redshift range is consistent
with $\gamma = 1.40$, thus we have a continuation of the trend 
seen at higher redshifts. 
Again, the transition to non-evolution probably occurs around $z=1$. 
More lines of sight are necessary to confirm our results.

\item

We detect no significant clustering of neither the weak nor 
the strong Lyman $\alpha$ lines on scales up to 10\,000 km\,s$^{-1}$,
so we are unable to confirm most of the previous studies reporting 
a weak clustering signal for $\Delta v \leq 150$ km\,s$^{-1}$. 
 
\end{enumerate}

\begin{acknowledgements}
  
This research has been supported by the Verbundforschung 
of the BMBF/DLR under Grant No.~50~OR~9911~1. We thank the 
anonymous referee for his very helpful report.

\end{acknowledgements}

%\bibliographystyle{aa}
%\bibliography{}
\end{document}